\documentstyle[emulateapj,apjfonts,psfig]{article}

\lefthead{Miller, Fabian, \& Miller}  
\righthead{M81 X-9 (Holmberg IX X-1)}

\received{Date}
\revised{Date}
\accepted{Date}

\journalid{vol}{date}
\articleid{1}{4}
\paperid{id}

\cpright{AAS}{1999}
\ccc{x}

\begin{document}

\title{Revealing a Cool Accretion Disk in the Ultraluminous X-ray Source M81 X-9 (Holmberg IX X-1): Evidence for an Intermediate Mass Black Hole}

\author{J.~M.~Miller\altaffilmark{1,2}, 
        A.~C.~Fabian\altaffilmark{3},
	M.~C.~Miller\altaffilmark{4}}

\altaffiltext{1}{Harvard-Smithsonian Center for Astrophysics, 60
	Garden Street, Cambridge, MA 02138, jmmiller@cfa.harvard.edu}
\altaffiltext{2}{NSF Astronomy and Astrophysics Fellow}
\altaffiltext{3}{Institute of Astronomy, University of Cambridge,
Madingley Road, Cambridge CB3 OHA, UK}
\altaffiltext{4}{Department of Astronomy, University of Maryland,
College Park, MD 20742}


\authoremail{jmmiller@cfa.harvard.edu}

\label{firstpage}

\begin{abstract}
We report the results of an analysis of two {\it XMM-Newton}/EPIC-pn
spectra of the bright ultraluminous X-ray source M81 X-9 (Holmberg IX
X-1), obtained in snapshot observations.  Soft thermal emission is
clearly revealed in spectra dominated by hard power-law components.
Depending on the model used, M81 X-9 was observed at a luminosity of
$L_{X} = 1.0-1.6 \times 10^{40}~{\rm erg}~{\rm s}^{1}$ (0.3--10.0
keV).  The variability previously observed in this source signals that
it is an accreting source which likely harbors a black hole.
Remarkably, accretion disk models for the soft thermal emission yield
very low inner disk temperatures ($kT = 0.17-0.29$~keV, including 90\%
confidence errors and variations between observations and disk
models), and improve the fit statistic over any single-component
continuum model at the $6\sigma$ level of confidence.  This
represents much stronger evidence for a cool disk than prior evidence
which combined spectra from different observatories, and the strongest
evidence of a cool disk in an ultraluminous X-ray source apart from
NGC 1313 X-1.  Like NGC 1313 X-1, scaling the temperatures measured in
M81 X-9 to those commonly seen in stellar-mass Galactic black holes at
their highest observed fluxes ($kT \simeq 1$~keV) may imply that M81
X-9 harbors a black hole with a mass on the order of
$10^{3}~M_{\odot}$; the measured disk component normalization and
broad-band luminosity imply black hole masses on the order of
$10^{2}~M_{\odot}$.  It is therefore possible that these sources
harbor $10^{3}~M_{\odot}$ black holes accreting at $L_{X} \simeq
0.1\times L_{Edd.}$.  While these results do not represent proof that
M81 X-9 harbors an intermediate-mass black hole, radio and optical
observations suggest that beaming and anisotropic emission from a
stellar-mass black hole are unlikely to account for the implied
luminosity.  We further argue that the strength of the hard emission
in these sources and well-established phenomena frequently observed in
stellar-mass black holes near to the Eddington limit suggest that
optically-thick photospheres are unlikely to be the origin of the cool
thermal emission in bright ultraluminous X-ray sources.  For
comparison to M81 X-9, we have also analyzed the previously
unpublished EPIC-pn spectrum of NGC 1313 X-1; cool disk emission is
again observed and refined spectral fit parameters and mass estimates
are reported.
\end{abstract}

\section{Introduction}
Ultraluminous X-ray sources (ULXs) are point-like X-ray sources in
nearby galaxies for which the implied luminosity exceeds the Eddington
limit ($L_{Edd.} = 1.3\times 10^{38}~ (M/M_{\odot})~{\rm erg}~{\rm
s}^{-1}$) for an isotropically-emitting black hole of $10~M_{\odot}$
(as per dynamically-constrained Galactic black holes; see McClintock
\& Remillard 2003).  A number of these sources were first identified
with {\it Einstein} (see, e.g., Fabbiano 1989), but the spatial
resolution and sensitivity of {\it Chandra} has shown
that most ULXs are likely point sources.  Short-term and
longer-term variability studies have shown that most ULXs must be
accreting sources, likely harboring black holes feeding from companion
stars in binary systems (for reviews, see Fabbiano \& White 2003;
Miller \& Colbert 2003).

Intermediate-mass black holes (IMBHs, $10^{2-5}~M_{\odot}$)
provide an attractive explanation for ULXs.  However, this interpretation
requires excellent evidence that alternative explanations are unlikely
to hold.  Viable models for ULXs which only require stellar-mass black
holes ($\simeq 10~M_{\odot}$) and which may explain at least the
lower-luminosity ULX sources include: anisotropic emission due to a
funnel-like inner disk geometry (King et al.\ 2001), strongly beamed
emission due to a line of sight coincident with a jet axis (Reynolds
1997; Kording, Falcke, \& Markoff 2002), and ``slim'' disks (Watarai,
Mizuno, \& Mineshige 2001) and/or radiation pressure-dominated
super-Eddington accretion disks (e.g., Begelman 2002).  

While some prior analyses of X-ray spectra (e.g., with {\it ASCA})
found some evidence for IMBHs, the sensitivity of these observations
only statistically required single-component spectral models (see
Colbert \& Mushotzky 1999, Makishima et al.\ 2000), which prevented
stronger conclusions.  The high effective area and sensitivity of {\it
XMM-Newton} are changing this observational constraint (Miller et al.\
2003, Strohmayer \& Mushotzky 2003).  Similarly, high-quality optical
observations of ULXs are now being made (e.g., Pakull \& Mirioni 2002)
which reveal relatively symmetric nebulae around some sources with
spectra indicating the ULX is acting on the local environment.  The
symmetry of these nebulae may indicate that these ULXs emit
isotropically, and that funneling does not hold.  Similarly, radio to
X-ray flux ratios may be used to constrain relativistic beaming
models.  Miller et al.\ (2003) analyzed {\it XMM-Newton}/EPIC-MOS
spectra of the ULXs NGC 1313 X-1 and X-2.  These ULX spectra were the
first to statistically require thermal disk and power-law continuum
components.  Remarkably, the disk temperatures measured in those
spectra are 5--10 times {\it lower} than those commonly measured in
stellar-mass black holes accreting at high rates.  Those temperatures,
the normalization of the disk components, and the very high
luminosities implied suggest that NGC 1313 X-1 and X-2 may harbor
IMBHs.  This suggestion is strengthened by symmetric optical nebulae
around X-1 and X-2 which suggest isotropic emission, and radio
observations which likely rule-out relativistic beaming (Miller et
al.\ 2003).  {\it XMM-Newton} observations of the brightest ULX in M82
reveal quasi-periodic oscillations and even an Fe~K$\alpha$ emission
line; the detection of these disk signatures strongly rules-out
beaming and likely also funneling, and also represents strong evidence
that the M82 ULX may harbor an IMBH (Strohmayer \& Mushotzky 2003).

The ULX M81 X-9 (actually in the dwarf galaxy Holmberg IX, but
hereafter referred to as M81 X-9) is an exceptionally bright ($L_{X} >
10^{40}~{\rm erg}~{\rm s}^{-1}$) source which is variable on the
timescale of weeks and months, with weak indications for a cool disk
component (La Parola et al.\ 2001).  The source position is roughly 2'
from the optical center of Ho~IX (Paturel \& Petit 2002).  
In order to better understand the nature of M81 X-9, we analyzed
spectra obtained in two short {\it XMM-Newton} observations available
in the public archive.  Herein, we report the results of fits to the
{\it XMM-Newton}/EPIC-pn spectra of M81 X-9.  For comparison, we also
analyzed the EPIC-pn spectrum of NGC 1313 X-1 (only the EPIC-MOS
spectra were analyzed previously).  Fits to these X-ray spectra
suggest that both sources may harbor IMBHs.  We discuss these results
within the context of other recent ULX observations, and prevalent
models for ULX accretion flow geometries.

\section{Observations and Data Reduction}
In this work, we consider the time-averaged {\it XMM-Newton}/EPIC-pn
spectra gathered in two observations of M81 X-9 and a single
observation of NGC 1313 X-1.  The data sets were obtained through the
{\it XMM-Newton} public data archive.  The EPIC-pn camera has a higher
effective area than the EPIC-MOS cameras, and drives the results of
any joint spectral analysis.  The results of an analysis of the
EPIC-MOS spectra of NGC 1313 X-1 and X-2 are reported in Miller et
al.\ (2003).  That analysis used the {\it XMM-Newton} reduction
and analysis suite SAS version 5.3.3, which failed to produce EPIC-pn
event lists for the observation of NGC 1313.  The reduction and
analysis reported in this work used SAS version 5.4.2, which
successfully produced an EPIC-pn event list for the NGC 1313
observation.  The results of fits to the EPIC-pn spectrum of NGC 1313
X-1 --- the stronger IMBH candidate of the two ULXs considered Miller
et al.\ (2003) --- are reported in this work.

M81 X-9 (Ho IX X-1) was observed on 10 April 2002 starting at 17:37:52
(UT), and again on 16 April 2002, starting at 17:33:15 (UT).  In both
cases, the camera was operated in ``PrimeFullWindow'' mode, with the
``thin'' optical blocking filter.  Application of the standard time
filtering resulted in net exposures of 7.8~ksec and 8.5~ksec,
respectively.  In both cases, events were extracted in a circle within
24'' of the source position ($09^{h}57^{m}54^{s}, 69^{\circ}03'46''$,
La Parola et al.\ 2001).  Due to the proximity of the source to the
edge of the CCD and the relatively few bright sources apparent in
surrounding region, background counts were extracted in an adjacent
circle with a radius of 24''.  NGC 1313 was observed by {\it
XMM-Newton} on 17 October 2000 starting at 03:59:23 (UT).  The camera
was operated in ``PrimeFullWindow'' mode with the ``medium'' optical
blocking filter.  Application of the standard time filtering resulted
in a net exposure of 29.3~ksec.  Events were extracted in a circle
within 24'' of the position reported by Miller et al.\ (2003).  A
background spectrum was extracted in an annulus between 24--30'',
again centered on the source position.

Each of the spectra were made by applying the selection criteria
described in the MPE ``cookbook''.  Briefly, these selections are as
follows: we set ``FLAG=0'' to reject bad pixels and events too close to
chip edges, event patterns 0--4 were allowed (these patterns
correspond to ``singles'' and ``doubles''), and the spectral channels
were grouped by a factor of 5.  The ``canned'' response files
appropriate to the CCD, patterns, and optical blocking filter were
used to fit the data.  Using the SAS tool ``epatplot'' and simulating
{\it XMM-Newton}/EPIC-pn spectra using the HEASARC tool ``PIMMS''
(based on prior reported fluxes and spectral parameters), we found
that the effects of photon pile-up are negligible in the spectra.
Prior to fitting, the source and background spectra were rebinned
(using ``grppha'' within LHEASOFT version 5.2) to require at least 15
counts per bin to ensure the validity of $\chi^{2}$ statistical
analysis.  Spectral fits were made using XSPEC version 11.2 (Arnaud \&
Dorman 2000).  All spectral fits were made in the 0.3--10.0~keV band.
All errors reported in this work are 90\% confidence errors, and were
evaluated while allowing all spectral model components to vary to take
account of dependencies between parameters and components.  For all
spectral fits, the standard F test was employed to calculate the
significance of adding model components.

\section{Analysis and Results}
We considered a number of continuum models commonly applied to
Galactic X-ray binary systems, modified by absorption in the
interstellar medium using the XSPEC model ``phabs''.  Although
somewhat technical, it is worth noting that fits to {\it
XMM-Newton}/RGS spectra of bright Galactic X-ray binaries show that
``phabs'' places the neutral oxygen absorption edge at an energy
slightly below that expected for neutral atomic oxygen (0.543~keV).
At CCD resolution, this can have the effect of creating a false
emission line feature at an energy just above that which ``phabs''
assumes for the oxygen edge.  Although different fixes for this
problem are possible, we adopted the following approach (in part
because it is easily reproducible).  Each spectrum was fit with a
model consisting of a simple power-law modified by ``phabs''; the edge
depth expected for the neutral oxygen edge based on this column and
assuming solar abundances was calculated; all continuum models then
fit to the spectra were modified by ``vphabs'' with the abundance of
oxygen set to zero, with an additional edge fixed at the calculated
depth added at 0.543~keV.  Regardless of the continuum model, this
absorption model proves to be a better fit to each spectrum in the
region of the oxygen edge, and removes the (false) apparent emission
line.

The results of all spectral fits are reported in Table 1.  The spectra
of M81 X-9 are shown in Figures 1, 2, 4, and 5, and the spectrum of
NGC 1313 X-1 is shown in Figures 3 and 6.  While a simple power-law
proves to be the best single-component continuum model in each case,
it is only an acceptable model for the latter observation of M81 X-9.
The canonical multi-color disk (MCD) blackbody model (Mitsuda et al.\
1984) --- an approximation to the Shakura \& Sunyaev (1973) disk model
--- is not an acceptable fit to any of the spectra when it is the only
continuum component.  Similarly, single-component thermal
bremsstrahlung models are not an acceptable fit to any of the spectra.
Though not listed in Table 1, single-component fits with any diffuse
thermal plasma model are not acceptable; this is expected given the
lack of emission lines in any spectrum.

Following Miller et al.\ (2003), we proceeded to consider
two-component continuum models: MCD plus power-law, ``diskpn'' plus
power-law, and ``MEKAL'' plus power-law.  While the MCD model is
convenient in that 1) it is a common component in published fits to
Galactic black hole binary spectra, 2) its temperature and
normalization allow scaling from which constraints on the mass of the
accretor in ULXs may be estimated, it is an approximation to the
Shakura \& Sunyaev (1973) disk in that it does not include the
zero-torque inner boundary condition.  The ``diskpn'' model is a disk
model that includes corrections (based on a pseudo-Newtonian
potential) for the temperature behavior near to the black hole
(Gierliski et al.\ 1999).  As with the MCD model, the temperature and
normalization parameters in this model permit scalings which can be
used to constrain the mass of the accretor in ULXs.  In addition, the
inner disk radius is also a free parameter in the diskpn model
($R_{in} = 6~R_{g}$ was fixed in all fits, where $R_{g} = GM/c^{2}$).
The ``MEKAL'' model describes diffuse thermal plasmas; the results of
fits with this model are not listed in Table 1 as the density
parameter of this model is totally unconstrained (likely, due to the
total lack of credible emission lines in the spectra), and because
``MEKAL'' plus power-law models prove to be significantly worse than
disk plus power-law models.  This is particularly clear in the case of
NGC 1313 X-1: the best-fit ``MEKAL'' plus power-law model gives
$\chi^{2}/\nu = 499.0/486$ (where $\nu$ is the number of degrees of
freedom); this is significantly worse than fits with a model
consisting of an MCD plus power-law components ($\chi^{2}/\nu =
459.8/487$) or diskpn plus power-law components ($\chi^{2}/\nu =
461.0/487$).  Finally, it is worth noting that archival observations
suggest that NGC 1313 X-1 and M81 X-9 are not diffuse sources, but are
constistent with point sources at {\it Chandra} resolution (after
accounting for pile-up).

Cool disk components are formally statistically required to fit the
first spectrum of M81 X-9 (the addition of a disk component is
significant at the $6\sigma$ level of confidence) and the spectrum of
NGC 1313 X-1 (the addition of a disk component is significant at more
than the $8\sigma$ level of confidence).  Figure 2 clearly shows that
there is also excess soft emission in the second spectrum of M81 X-9.
It was shown previously that soft excesses of this kind are robust
against under-abundances in the local absorbing material (Miller et
al.\ 2003a), and that result holds for these spectra as well.
When a disk component is included in the model, the parameters of the
disk component can be constrained and the addition gives an
improvement which is significant at nearly the $7\sigma$ level.
Remarkably, the disk components in fits to M81 X-9 yield very low
temperatures (ranging between $kT = 0.17-0.29$~keV, including 90\%
confidence errors, and considering each of the two models fit to the
two observations of M81 X-9).  Inner disk temperatures ranging between
$kT = 0.21-0.25$~keV are measured for the disk component in the
spectrum of NGC 1313 X-1.  The power-law indices measured are harder
than those commonly measured in the ``very high state'' in
stellar-mass Galactic black holes, but softer than those commonly
measured in the ``low/hard'' state.

Estimates for the mass of the black holes in M81 X-9 and NGC 1313 X-1
are given in Table 1, based on three independent methods: 1) scaling
the inner disk temperature to values commonly measured in Galactic
black holes, 2) direct estimates from the normalization of the disk
components, and 3) scaling from the Eddington limit equation.  

At the highest mass accretion rates observed in stellar-mass black
holes, inner disk temperatures near to or above $kT = 1$~keV are often
observed (for a review, see McClintock \& Remillard 2003; for a recent
example, see Park et al.\ 2003).  In some cases, however, the inner
disk temperatures can be as high as $kT \simeq 2$~keV (for examples
and a comparison to ULXs, see Makishima et al.\ 2000).  Using the fact
that in blackbody disk models, $T \propto M^{-1/4}$, it is possible to
scale temperature to mass via the relation $(M_{ULX}/M_{10~M_{\odot}})
\propto (kT_{10~M_{\odot}}/kT_{ULX})^{4}$.  Assuming that M81 X-9 and
NGC 1313 X-1 are accreting at or near to the Eddington limit (the
regime in which hot disks are observed in stellar-mass Galactic black
holes), this scaling suggests that these ULXs harbor black holes with
masses of $few \times 10^{3}~M_{\odot}$ (see Table 1).  Clearly, this
scaling is very sensitive to what temperature is assumed to be typical
of stellar-mass Galactic black holes accreting at high rates.
Assuming $kT =0.5$~keV to be typical gives masses of $few \times
10^{2}~M_{\odot}$ for M81 X-1 and NGC 1313 X-1; however, assuming $kT
= 1.0$~keV to be typical for stellar-mass black holes accreting at or
near to the Eddington limit has a much stronger observational basis.

The MCD and diskpn normalizations also allow estimates of the black
hole masses.  In this case, some care is required because it has been
noted that the color temperature measured using multi-color disk
models may differ from the effective temperature (Shimura \& Takahara
1995; Merloni, Fabian, \& Ross 2000).  Whereas corrections divide-out
in scaling the temperature, they must be considered in scaling masses
from the disk component normalization.  The MCD normalization gives
the inner disk extent in kilometers, which can be converted into a
black hole mass.  Following recent related work (Miller et al.\ 2003;
see also Sobczak et al.\ 2000 and Makishima et al.\ 2000), we take
$R_{in,corr} = \eta f^{2} R_{obs.}$ for the MCD model (where $f=1.7$
is the spectral hardening factor, and $\eta=0.63$ is valid for
$i<70^{\circ}$ and accounts for the difference between the innermost
radius and the radius of peak temperature; note also that in all disk
normalization scalings we assumed $i=0$ and that any higher
inclination gives a {\it higher} mass).  For a Schwarzschild black
hole, $R_{in} = 6~R_{g} = 8.85~{\rm km} / M_{\odot}$ (where $R_{g} =
GM/c^{2}$).  Therefore, we can use the MCD model to obtain mass
estimates via the relation: $M_{ULX} = \eta f^{2} (K/cos[i])^{1/2}
\times (d / 10~{\rm kpc}) \times (8.85 {\rm km})^{-1}$ (where $K$ is
the MCD normalization, $i$ is the inclination, and $d$ is the source
distance).  Assuming distances of $d=3.4$~Mpc for Ho~IX (Georgiev
1991, Hill et al.\ 1993) and $d=3.74$~Mpc for NGC 1313 (Tully 1998),
the MCD component normalization suggests that M81 X-9 and NGC 1313 X-1
may harbor black holes with masses of $few \times 10^{2}~M_{\odot}$ (see
Table 1).  The normalization of the diskpn component explicitly
includes the mass and the color correction factor $f$.  In estimating
masses with the diskpn model, $f=1.7$ was again assumed, as per the
MCD model.  The diskpn normalizations measured for M81 X-9 and NGC
1313 X-1 again suggest that these ULXs may harbor black holes with
masses of $few \times 100~M_{\odot}$ (see Table 1).

The mass of the putative black holes powering M81 X-9 and NGC 1313 X-1
can also be estimated by scaling the implied source luminosities to
the Eddington limit for isotropically emitting accretion-powered
sources ($L_{Edd.} = 1.3 \times 10^{38}~ (M/M_{\odot})~ {\rm erg}~{\rm
s}^{-1} / M_{\odot}$; Frank, King, \& Raine 2002).  To make this
scaling, the unabsorbed 0.3--10.0~keV flux was extrapolated to the
0.05--100.0~keV band, as the latter is more representative of a
bolometric luminosity.  This scaling again implies that M81 X-9 and
NGC 1313 X-1 may harbor black holes with masses of $few \times
10^{2}~M_{\odot}$ (see Table 1).  It is interesting to note that
scaling the ULX inner disk temperatures to $kT = 1.0$~keV implies
masses on the order of $10^{3}~M_{\odot}$ while the disk
normalizations and overall luminosities imply masses on the order of
$10^{2}~M_{\odot}$.  This may imply that these ULXs are
$10^{3}~M_{\odot}$ black holes seen at $L_{X} \simeq 0.1\times
L_{Edd.}$.

Finally, we note that while the results obtained here for NGC 1313 X-1
broadly agree with those previously reported in Miller et al.\ (2003),
they do not agree exactly.  The differences can be attributed to
several factors, including: the more conservative fitting range used
in this analysis (0.3--10.0~keV) relative to the prior analysis, the
higher effective area, sensitivity, and energy resolution of the
EPIC-pn camera (used in this analysis) relative to the EPIC-MOS
cameras (used in the prior analysis), and the improved interstellar
absorption model used in this analysis.  The most important difference
is that the disk temperatures measured in this analysis are 0.08~keV
higher than in the previous analysis ($kT = 0.23$~keV versus $kT =
0.15$~keV), resulting in new mass estimates which are five times lower
than prior estimates using inner disk temperatures.

\section{Discussion}
We have investigated the nature of the ULX M81 X-9 (Holmberg IX X-1)
by analyzing two {\it XMM-Newton}/EPIC-pn spectra of the source.  The
most important result of this work is that cool thermal continuum
emission is unambiguously required for the first time.
Optically-thick disk components yield improvements in the fit
statistic which are significant at more than the $6\sigma$ level of
confidence.  This represents the clearest evidence for a cool
accretion disk in a ULX apart from NGC 1313 X-1 (Miller et al.\ 2003;
see Table 1).  

Remarkably, the inner disk temperatures measured ($kT =
0.17-0.29$~keV, including 90\% confidence errors on two disk models
fit to the two observations) are well below those commonly measured in
stellar-mass black holes ($kT \simeq 1$~keV, see McClintock \&
Remillard for a review).  Scaling the temperatures measured implies
that M81 X-9 may harbor a black hole with a mass on the order of
$10^{3}~M_{\odot}$, and scaling the disk component normalizations
implies masses on the order of $10^{2}~M_{\odot}$.  It is unclear
which scaling method is superior, but both scalings suggest that M81
X-9 may harbor an IMBH.  Both scaling methods can be correct if the
source harbors a black hole with a mass on the order of
$10^{3}~M_{\odot}$ which is observed at $L_{X} \simeq 0.1\times
L_{Edd.}$.  For comparison, we also report results from an analysis of
the EPIC-pn spectrum of NGC 1313 X-1, which may also harbor an IMBH
(Miller et al.\ 2003).  The spectra of these sources bear striking
similarities (see Table 1).

The long-term and short-term flux variability of M81 X-9 demonstrated
clearly by La Parola et al.\ (2001) establishes that M81 X-9 is an
accreting source.  While it is natural to ascribe cool thermal
emission to an optically-thick accretion disk in accreting systems,
given the implications of these cool disks it is especially important
to understand why disk emission is the most viable explanation.

Relativistic beaming --- which might prevent detection of disk
signatures --- is not likely at work in M81 X-9.  First, radio
observations of Ho IX by Bash \& Kaufman (1986) find a flux density of
at most 1~mJy at 20~cm at the position of the ULX, corresponding to a
luminosity of $\simeq 2\times 10^{34}~{\rm erg}~{\rm s}$.  If the
source was active at its present level during that time (as suggested
by the long-term behavior reported by La Parola et al.\ 2001), this
radio luminosity gives a radio to X-ray luminosity ratio of $\simeq
2\times 10^{-6}$.  This is at odds with a relativistic beaming
scenario, as beaming should create flat spectra (see, e.g., Fossati et
al.\ 1998).  Second, the peak radio to X-ray flux ratios observed in
Galactic X-ray binary systems are always below $2.3\times 10^{-5}$
(Fender \& Kuulkers 2001; Barth, Ho, \& Sargent 2003).  The fact that
the radio to X-ray flux ratio in M81 X-9 is likely an order of
magnitude below the highest ratio seen in Galactic systems suggests
that M81 X-9 is not likely to be a stellar-mass black hole with
relativistically-beamed emission coincident with our line of sight to
the source.

Slightly anisotropic X-ray emission --- perhaps due to a funnel-like
geometry in the inner disk --- might create a very hot inner disk,
especially if the source of hard X-ray emission is central (e.g.,
through Comptonization).  Clearly, the spectra we have analyzed
rule-out hot disk signatures.  Pakull \& Mirioni (2002) have reported
optical nebulae around a few ULXs, including NGC 1313 X-1 and M81 X-9.
The line ratios observed in these nebulae suggest that X-ray
photoionization may be important, and hint that some ULXs may act on
their local ISM.  The symmetry of the nebula around NGC 1313 X-1
suggests that the source emits isotropically.  While the nebula imaged
around M81 X-9 is better described by an ellipse than a circle, the
ratio of the axes is likely less than 2:1, and certainly inconsistent
with a ratio of 10:1 (``funneled'' scenarios require an anisotropy
parameter of approximately 10; see King et al.\ 2001).  Thus, present
data suggests that anisotropic emission is not the best explanation
for the high flux observed from M81 X-9.

Radiation pressure-dominated accretion disks --- which might be able to
produce super-Eddington fluxes through small-scale photon bubble
instabilities (Begelman 2002) --- might be expected to have especially
high temperatures (perhaps similar to the ``slim'' disk solution
described by Watarai, Mizuno, \& Mineshige 2001).  As the spectra of
M81 X-9 and NGC 1313 X-1 are inconsistent with hot disk components, it
is not likely that such models describe these sources.

Finally, it has recently been suggested that soft thermal emission in
the brightest ULXs may be due to an optically-thick, outflowing
photosphere originating at $100~R_{Schw.}$ around a stellar-mass black
hole accreting at a super-Eddington rate (e.g., King 2003).  This
model is inconsistent with the relative importance of hard X-ray
emission in sources like M81 X-9 and NGC 1313 X-1.  It is
also inconsistent with a number of long-established properties
observed in stellar-mass Galactic black holes accreting near to, at,
or slightly in excess of their isotropic Eddington luminosities.  We
will briefly consider these issues here.

The mechanical power in a photosphere outflowing at velocity $v$
should scale with the luminosity radiated by the photosphere as
$L_{mech.} \simeq L_{rad}\times (v/c)$, and that the photospheric
radius is given by $R_{phot} \simeq (c/v)^{2} \times R_{Schw.}$ (King
2003).  Hard X-ray emission in accreting sources is usually tied to
regions deep within the gravitational potential, as this is a
convenient energy reservoir.  However, the innermost accretion regime
would be blocked from view by an optically-thick photosphere.  The
observed hard component in such a scenario must be generated in shocks
above the photosphere, which means that the strength of the hard
component must be driven by the outflow: $L_{hard} \simeq L_{mech}$.
Measurements indicate that the hard components in M81 X-9 and NGC 1313
X-1 are 3--4 times stronger than the soft (putatively photospheric)
components in each spectrum (see Table 1).  Even if we only assume
that $L_{hard} \simeq L_{soft} \simeq L_{rad}$, the only way $L_{mech}
\simeq L_{rad}\times (v/c)$ holds is if $(v/c) \simeq 1$, which
implies that $R_{phot} \simeq (c/v)^{2}\times R_{Schw.} \simeq
1~R_{Schw}$.

Moreover, even at the highest implied accretion rates, a number of
spectral and variability properties observed in stellar-mass
Galactic black holes would be screened if an optically-thick
photosphere developed at $R_{phot} \simeq 100~R_{Schw.}$.  First, hot
($kT \simeq 1$~keV) thermal spectral components are observed in every
source at peak X-ray flux; this emission can only be associated with
the inner disk as a radially distant photosphere should be much
cooler.  Second, QPOs -- whether high frequency ($few \times 100$~Hz)
or low frequency ($few \times 1$~Hz) -- are almost certainly tied to the
disk, and likely represent Keplerian orbital frequencies and/or
coordinate resonance frequencies.  QPOs are seen over a factor of
$10^{3}$ in flux in stellar-mass Galactic black holes but are
preferentially seen at high fluxes.  Most importantly, QPOs are a
higher fraction of the rms noise at high energies -- QPOs are
intimately tied to hard X-ray emission.  It is very unlikely that such
periodicities originate in shocks above an optically-thick
photosphere.  Third, broad, relativistic Fe~K$\alpha$ emission lines
are also observed in stellar-mass Galactic black holes over a factor
of $10^{3}$ in flux, but preferentially at the very highest fluxes
(e.g., in the very high state).  These lines are certainly tied to
hard X-ray emission (see, e.g., Zdziarski, Lubinski, \& Smith 1999),
and have been observed to vary at frequencies as high as 6~Hz
(Gilfanov, Churazov, \& Revnivtsev 2000).  These facts -- and their
smeared shape (very likely due to the extreme Doppler shifts
and gravitational red-shift at the inner disk) -- tie broad
Fe~K$\alpha$ emission lines to the innermost accetion flow.  The
reader is directed to McClintock \& Remillard (2003), and references
therein, for a discussion of the stellar-mass black hole properties
noted here.

It is worth noting that partial covering models for NGC 1313 X-1 and
M81 X-9 cannot be ruled-out statistically in the limited energy range
considered.  Models consisting of a strong absorber ($N_{H} \simeq
10^{23}~{\rm cm}^{-2}$) covering 50--70\% of a steep power-law source
($\Gamma \simeq 2.7$) can fit the data as well as our disk plus
power-law models.  There are several reasons, however, why partial
covering is likely an unphysical model for these sources.  First,
partial covering is only required in Galactic black hole sources
during X-ray "dips" (e.g., 4U 1630$-$472 and GRO~J1655$-$40, Kuulkers
et al.\ 1998).  When partial covering is actually required to fit the
spectra of Galactic black holes and AGN, fits with simple models
require very strong neutral Fe~K absorption edges: Ueda et al.\ (1998)
find an edge with $\tau = 3.3$ in GRO~J1655$-$40, and Boller et al.\
(2002) report an edge with $\tau = 1.8$ in 1H~0707$-$495.  In
contrast, X-ray dips have never been observed in NGC 1313 X-1 or M81
X-9, and the 95\% confidence upper limits on the strength of a neutral
Fe~K edges in the spectra of NGC 1313 X-1 and M81 X-9 are $\tau \leq
0.4$ and $\tau \leq 0.2$, respectively.  Next, it has been shown that
while partial covering models can provide acceptable fits to AGN data
below 10~keV, simultaneous high energy spectra strongly reject such
models (see, e.g., Reynolds et al.\ 2004).  Finally, the above
discussion of photospheric models shows that an alternative geometry
wherein partial covering might arise naturally is very unlikely to
describe sources like NGC 1313 X-1 and M81 X-9.

Thus, at present, the most viable explanation for the soft thermal
excess observed in sources like M81 X-9 and NGC 1313 X-1 is emission
from a cool accretion disk around an IMBH.  Recent observations of
bright ULXs with {\it XMM-Newton} and {\it Chandra} have revealed cool
thermal (likely disk) components in a number of ULXs, including (but
not limited to) NGC 1313 X-1 and X-2 (Miller et al.\ 2003), NGC 5408
X-1 (Kaaret et al.\ 2003), NGC 4038/4039 X-37 (Miller et al.\ 2003b),
and M74 X-1 (Krauss et al.\ 2003).  The brightest ULX in M82 may or
may not have a cool disk; the low energy portion of the spectrum from
that ULX has proved difficult to measure due to photon pile-up in CCD
spectrometers (with {\it Chandra}) and diffuse emission in the crowded
field (with {\it XMM-Newton}).  However, beaming is confidently
ruled-out in this source through the observation of QPOs and an
Fe~K$\alpha$ emission line (Strohmayer \& Mushotzky 2003), and this source
is an excellent IMBH candidate.  At most, these observations suggest
that sources at the upper-end of the ULX luminosity distribution {\it
may} harbor IMBHs.  Many ULXs at the lower end of the luminosity
distribution may be stellar-mass black holes.  As observations of ULXs
continue to be made (especially, multi-wavelength observations), it is
likely that sub-classes of beamed, funneled, and IMBH sources will be
distinguished.

If M81 X-9 harbors an IMBH, we can speculate about how such a black
hole was made.  This speculation is particularly interesting because
Ho IX is a dwarf galaxy.  Madau \& Rees (2001) have proposed IMBHs
might have been created by the death of extremely massive,
low-metallicity Population III stars.  While most IMBHs created in
this way may have been dragged to galactic centers by dynamical
friction, a fraction might be visible today as ULXs if they have
captured a donor star.  If M81 X-9 has a mass on the order of
$10^{2}~M_{\odot}$ and this scenario can hold in the case of dwarf
galaxies, M81 X-9 could be a Population III remnant.  If M81 X-9 has a
mass on the order of $10^{3}~M_{\odot}$, it is more likely that it
grew through mergers, either in a young cluster (e.g., Ebisuzaki et
al.\ 2001) or in a globular cluster (Miller \& Hamilton 2002).
Certainly, Miller (1995) notes that the vicinity of M81 X-9 may
contain a number of massive, young stars, and may show evidence for a
recent supernova history.

J. M. M. acknowledges support from the NSF through its Astronomy and
Astrophysics Postdoctoral Fellowship program, and useful discussions
with A. Kong and P. Slane.  M. C. M. was supported in part by NSF
grant AST 00-98436, and by NASA grant NAG 5-13229.  This work is based
on observations obtained with {\it XMM-Newton}, an ESA mission with
instruments and contributions directly funded by ESA member states and
the US (NASA).  This work has made use of the tools and services
available through HEASARC online service, which is operated by GSFC
for NASA.  The authors note for the curious that J. M. Miller (Miller
et al.\ 2003a, 2003b), M. C. Miller (Miller \& Hamilton 2002, Miller
\& Colbert 2003), and B. F. Miller (Miller 1995) are indeed distinct,
unrelated people.

\begin{table}[h]
\caption{Spectral Fit Parameters}
\begin{footnotesize}
\begin{center}
\begin{tabular}{lllll}
\multicolumn{2}{l}{Model/Parameter} & M81 X-9 Obs.\ 1 & M81 X-9 Obs.\ 2 & NGC 1313 X-1\\
\tableline

\multicolumn{2}{l}{power-law} & ~ & ~ & ~ \\
\multicolumn{2}{l}{$N_{H}~(10^{21}~{cm}^{-2})$} & $2.1(1)$ & $2.5(2)$ & $2.4(2)$ \\
\multicolumn{2}{l}{$\Gamma$} & $1.93(4)$ & $2.00(4)$ & $2.04(5)$ \\
\multicolumn{2}{l}{Norm. ($10^{-3}$)} & $1.24(4)$ & $1.57(5)$ & $0.69(3)$ \\
\multicolumn{2}{l}{$\chi^{2}/dof$} & 460.7/443 & 571.7/579 & 555.7/489 \\
\tableline

\multicolumn{2}{l}{MCD} & ~ & ~ & ~ \\
\multicolumn{2}{l}{$N_{H}~(10^{21}~{cm}^{-2})$} & $2.1$ & $2.5$ & $2.4$ \\
\multicolumn{2}{l}{$kT$~(keV)} & $0.91(5)$ & $1.63(4)$ & $0.78(5)$  \\
\multicolumn{2}{l}{Norm. ($10^{-1}$)} & $3.5(2)$ & $0.40(2)$ & $3.2(2)$ \\
\multicolumn{2}{l}{$\chi^{2}/dof$} & 1823/444 & 1421/579 & 1919/490 \\
\tableline

\multicolumn{2}{l}{bremsstrahlung} & ~ & ~ & ~ \\
\multicolumn{2}{l}{$N_{H}~(10^{21}~{cm}^{-2})$} & $2.1$ & $2.5$ & $2.4$ \\
\multicolumn{2}{l}{$kT$~(keV)} & $31(1)$ & $43(1)$ & 4.6(1) \\
\multicolumn{2}{l}{Norm. ($10^{-3}$)} & $1.04(2)$ & $1.22(2)$ & $0.70(2)$ \\
\multicolumn{2}{l}{$\chi^{2}/dof$} & 1510/443 & 2435/579 & 773/489 \\
\tableline

\multicolumn{2}{l}{MCD $+$ power-law} & ~ & ~ & ~ \\
\multicolumn{2}{l}{$N_{H}~(10^{21}~{cm}^{-2})$} & $2.3(3)$ & $2.9(3)$ & $3.1(3)$ \\
\multicolumn{2}{l}{$kT$~(keV)} & $0.26^{+0.02}_{-0.05}$ & $0.21(4)$ & $0.23(2)$ \\
\multicolumn{2}{l}{Norm.} & $20^{+20}_{-10}$ & $60^{+70}_{-40}$ & $28(5)$ \\
\multicolumn{2}{l}{$\Gamma$} & $1.73(8)$ & $1.86(6)$ & $1.76(7)$ \\
\multicolumn{2}{l}{Norm. ($10^{-3}$)} & $0.96(9)$  & $1.34^{+0.09}_{-0.05}$ & $0.49(4)$ \\
\multicolumn{2}{l}{$\chi^{2}/dof$} & 419.0/441 & 524.3/577 & 459.8/487 \\
\multicolumn{2}{l}{disk significance} & $6.1\sigma$ & $6.8\sigma$ & $>8\sigma$ \\
\multicolumn{2}{l}{$F~(10^{-12}~erg~cm^{-2}~s^{-1})$} & $8^{+2}_{-1}$ & $10^{+2}_{-1}$ & $4.3(4)$ \\
\multicolumn{2}{l}{$F_{power-law}/F_{total}$} & 0.85  & 0.84 & 0.74 \\
\multicolumn{2}{l}{$L_{0.3-10}~(10^{40}~erg~{s}^{-1})$} & $1.1^{+0.3}_{-0.1}$ & $1.3^{+0.3}_{-0.2}$ & $0.6(1)$ \\
\multicolumn{2}{l}{$L_{0.05-100}~(10^{40}~erg~{s}^{-1})$} & $2.7^{+0.7}_{-0.3}$ & $2.9^{+0.6}_{-0.3}$ & $1.4^{+0.2}_{-0.2}$ \\
\multicolumn{2}{l}{$M_{kT = 1.0}~(M_{\odot})$} & $2000^{+3000}_{-1000}$ & $5000^{+7000}_{-2000}$ & $4000^{+2000}_{-1000}$ \\
\multicolumn{2}{l}{$M_{kT = 0.5}~(M_{\odot})$} & $140^{+180}_{-40}$ & $320^{+430}_{-160}$ & $220^{+100}_{-60}$ \\
\multicolumn{2}{l}{$M_{Norm.}~(M_{\odot})$} & $330^{+140}_{-100}$ & $570^{+270}_{-240}$ & $400^{+40}_{-40}$ \\
\multicolumn{2}{l}{$M_{L_{X}/L_{Edd.}}~(M_{\odot})$} & $200^{+50}_{-30}$ & $220^{+40}_{-20}$ & $110^{+10}_{-10}$ \\
\tableline

\multicolumn{2}{l}{diskpn $+$ power-law} & ~ & ~ & ~ \\
\multicolumn{2}{l}{$N_{H}~(10^{21}~{cm}^{-2})$} & $2.3(3)$ & $2.8(3)$ & $3.0(3)$ \\
\multicolumn{2}{l}{$kT$~(keV)} & $0.24(5)$ & $0.21(4)$ & $0.22(3)$ \\
\multicolumn{2}{l}{Norm. ($10^{-4}$)} & $4^{+16}_{-2}$ & $7^{+30}_{-3}$ & $0.5^{+0.5}_{-0.2}$ \\
\multicolumn{2}{l}{$\Gamma$} & $1.74^{+0.07}_{-0.04}$ & $1.85(6)$ & $1.76(7)$ \\
\multicolumn{2}{l}{Norm. ($10^{-3}$)} & $1.0(1)$  & $1.3^{+0.2}_{-0.1}$ & $0.49(5)$ \\
\multicolumn{2}{l}{$\chi^{2}/dof$} & 418.8/441 & 524.9/577 & 461.0/487 \\
\multicolumn{2}{l}{disk significance} & $5.9\sigma$ & $6.7\sigma$ & $>8\sigma$ \\
\multicolumn{2}{l}{$F~(10^{-12}~erg~cm^{-2}~s^{-1})$} & $8^{+2}_{-1}$ & $10^{+2}_{-1}$ & $4.3(4)$ \\
\multicolumn{2}{l}{$F_{power-law}/F_{total}$} & 0.85 & 0.85 & 0.74 \\
\multicolumn{2}{l}{$L_{0.3-10}~(10^{40}~erg~{s}^{-1})$} & $1.1^{+0.3}_{-0.1}$ & $1.3^{+0.3}_{-0.2}$ & $0.6(1)$ \\
\multicolumn{2}{l}{$L_{0.05-100}~(10^{40}~erg~{s}^{-1})$} & $2.7^{+0.7}_{-0.3}$ & $2.9^{+0.6}_{-0.3}$ & $1.4^{+0.2}_{-0.2}$ \\
\multicolumn{2}{l}{$M_{kT = 1.0}~(M_{\odot})$} & $3000^{+5000}_{-2000}$ & $5000^{+7000}_{-2000}$ & $4000^{+3000}_{-2000}$ \\
\multicolumn{2}{l}{$M_{kT = 0.5}~(M_{\odot})$} & $200^{+300}_{-100}$ & $300^{+400}_{-100}$ & $300^{+200}_{-100}$  \\
\multicolumn{2}{l}{$M_{Norm.}~(M_{\odot})$} & $330^{+20}_{-20}$ & $380^{+20}_{-20}$ & $240^{+10}_{-20}$  \\
\multicolumn{2}{l}{$M_{L_{X}/L_{Edd.}}~(M_{\odot})$} & $200^{+50}_{-30}$ & $220^{+40}_{-20}$ & $110^{+10}_{-10}$ \\
\tableline

\tableline
\end{tabular}
\vspace*{\baselineskip}~\\ \end{center} 
\tablecomments{The results of fitting simple models to the EPIC-pn
  spectra of M81 X-9 and NGC 1313 X-1.  The XSPEC model ``vphabs''was
  used to measure the equivalent neutral hydrogen column density along
  the line of sight.  All errors reported above are 90\% confidence
  errors.  The significance of adding disk components was calculated
  using the standard F test.  Errors in parentheses are the error in
  the last digit.  Fluxes are ``unabsorbed'' fluxes, and the reported
  luminosities assume a distance of 3.4~Mpc for Ho~IX and 3.7~Mpc for
  NGC 1313.  Mass estimates based on scaling the inner disk
  temperatures to those commonly measured in $10~M_{\odot}$ black
  holes accreting at high rates assume $T \propto M_{BH}^{-1/4}$.
  Scalings are given assuming typical inner disk temperatures of $kT =
  1.0$~keV and $kT = 0.5$~keV, although the literature favors the
  former.  Please see the text for details on mass estimates obtained
  through the disk component normalizations.  Mass estimates based on
  the Eddington luminosity make use of the 0.05--100~keV luminosity
  (extrapolated from the 0.3--10.0~keV flux) and assume $L_{Edd} =
  1.3\times 10^{38}~ (M/M_{\odot})~{\rm erg/s}$).}
\vspace{-1.0\baselineskip}
\end{footnotesize}
\end{table}

\clearpage

\centerline{~\psfig{file=f1.ps,width=3.2in,angle=-90}~}
\figcaption[h]{\footnotesize The ratio of the EPIC-pn spectrum of M81
X-9 obtained in the first observation to a model consisting of only
power-law continuum fit in the 3.0--10.0~keV band.  The soft continuum
excess is readily apparent at low energy.  The spectrum shown has been
rebinned for visual clarity.}
\medskip

\centerline{~\psfig{file=f2.ps,width=3.2in,angle=-90}~}
\figcaption[h]{\footnotesize The ratio of the EPIC-pn spectrum of M81
X-9 obtained in the second observation to a model consisting of only
power-law continuum fit in the 3.0--10.0~keV band.  The soft continuum
excess is readily apparent at low energy.  The spectrum shown has been
rebinned for visual clarity.}
\medskip

\centerline{~\psfig{file=f3.ps,width=3.2in,angle=-90}~}
\figcaption[h]{\footnotesize The ratio of the EPIC-pn spectrum of NGC
1313 X-1 obtained in the second observation to a model consisting of
only power-law continuum fit in the 3.0--10.0~keV band.  The soft continuum
excess is readily apparent at low energy.  The spectrum shown has been
rebinned for visual clarity.}
\medskip

\centerline{~\psfig{file=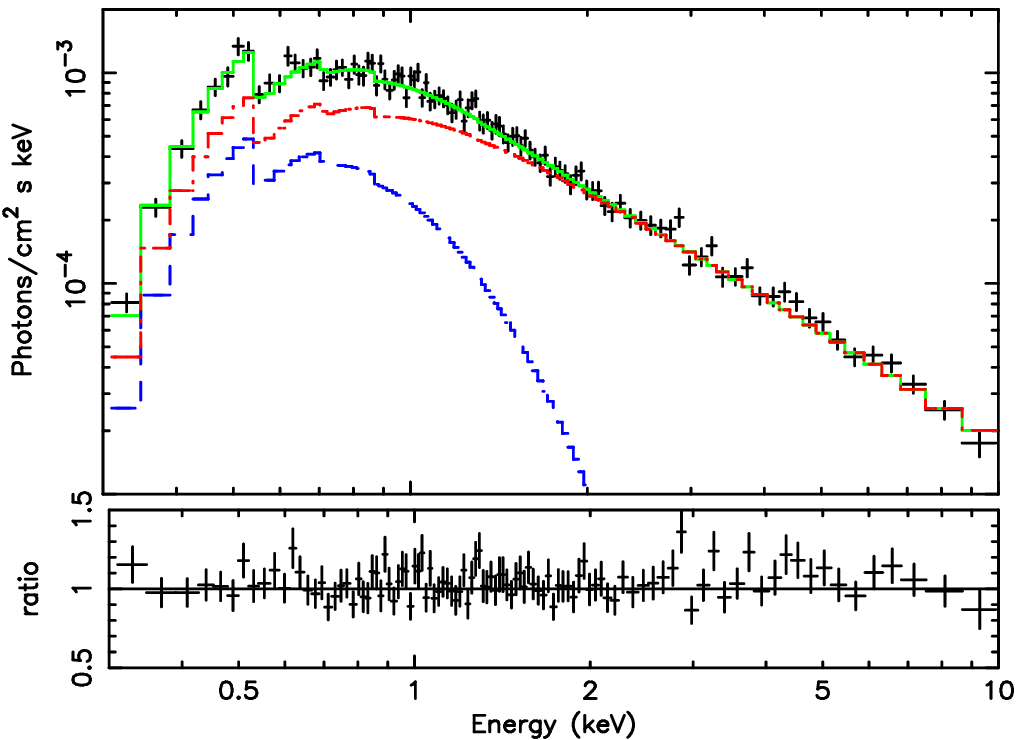,width=3.2in}~}
\figcaption[h]{\footnotesize The EPIC-pn spectrum of M81 X-9 obtained
with the first observation of the source.  The spectrum has been fit
with a model consisting of MCD (blue) and power-law (red) continuum
components, and the data/model ratio is shown below.  The shape of the
disk component and ratio plot are nearly identical if the ``diskpn''
disk model is fit instead of the MCD model.  See Table 1 for details
of the spectral model.}
\medskip

\centerline{~\psfig{file=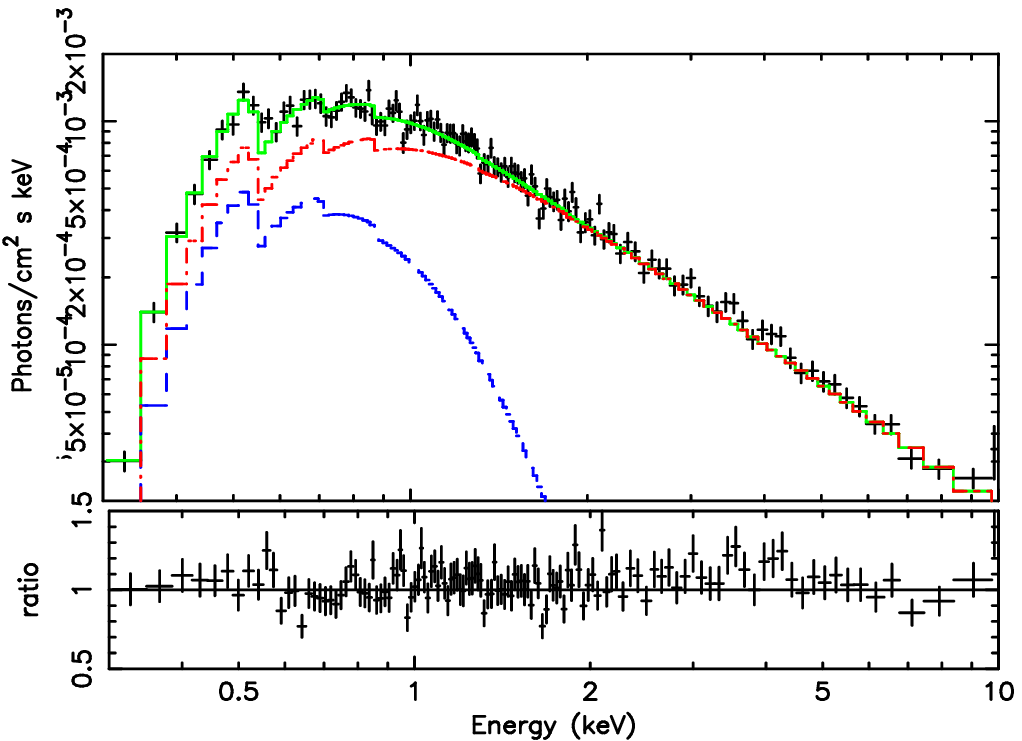,width=3.2in}~}
\figcaption[h]{\footnotesize The EPIC-pn spectrum of M81 X-9 obtained
with the second observation of the source.  The spectrum has been fit
with a model consisting of MCD (blue) and power-law (red) continuum
components, and the data/model ratio is shown below.  The shape of the
disk component and ratio plot are nearly identical if the ``diskpn''
disk model is fit instead of the MCD model.  See Table 1 for details
of the spectral model.}
\medskip

\centerline{~\psfig{file=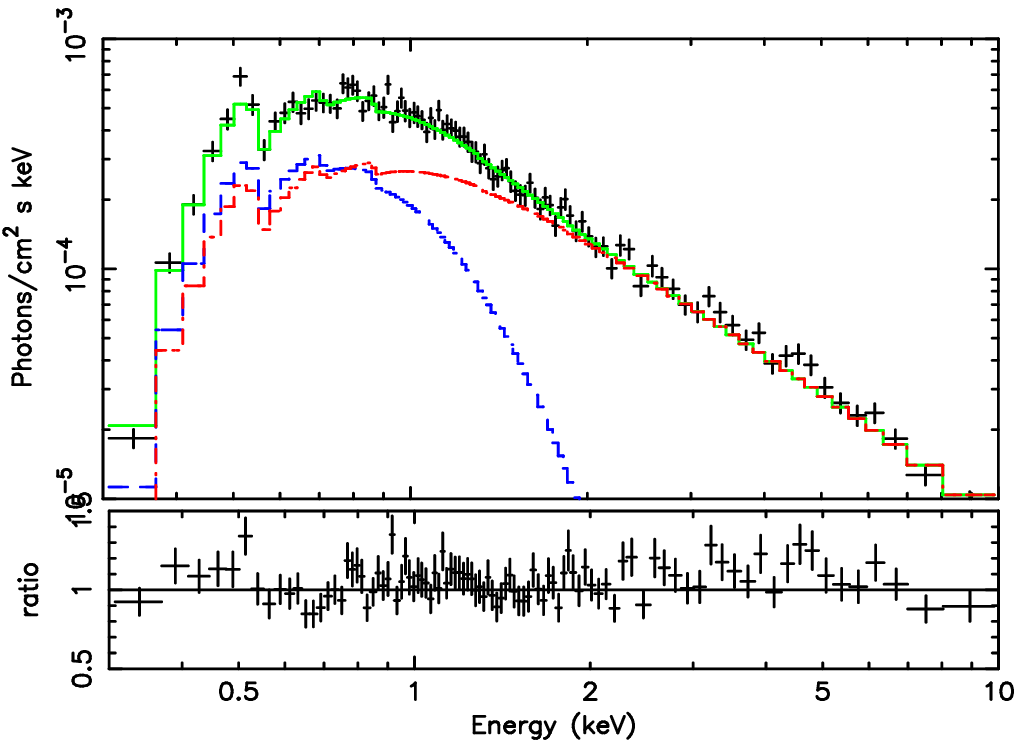,width=3.2in}~}
\figcaption[h]{\footnotesize The EPIC-pn spectrum of NGC 1313 X-1 obtained
with the first observation of the source.  The spectrum has been fit
with a model consisting of MCD (blue) and power-law (red) continuum
components, and the data/model ratio is shown below.  The shape of the
disk component and ratio plot are nearly identical if the ``diskpn''
disk model is fit instead of the MCD model.  See Table 1 for details
of the spectral model.}
\medskip

\end{document}